\documentclass[twocolumn,10pt,aps,amsmath,amssymb,pra]{revtex4-1}

\usepackage{graphicx}

\newcommand{\ket}[1]{{\vert #1 \rangle}}
\newcommand{\bra}[1]{{\langle #1 \vert}}
\newcommand{\braket}[2]{{\langle #1 \vert #2 \rangle}}

\newcommand{\Tr}{{\mathrm{Tr}}}

\begin{document}

\title{Time evolution of projected entangled pair states in the single layer picture}

\author{Iztok Pi\v{z}orn}
\author{Ling Wang}
\author{Frank Verstraete}
\affiliation{Faculty of Physics, University of Vienna, Boltzmanngasse 5, A-1090 Wien}

\date{March 11, 2011}

\begin{abstract}
We propose an efficient algorithm for simulating quantum many-body systems in two spatial
dimensions using projected entangled pair states. This is done by approximating
the environment, arising in the context of updating tensors in the process of
time evolution, using a single-layered tensor network structure. This
significantly reduces the computational costs and allows simulations in a
larger submanifold of the Hilbert space as bounded by the bond dimension of
the tensor network. We present numerical evidence for stability of the method
on an antiferromagnetic isotropic Heisenberg model where good agreement is
found with the available accurate results.
\end{abstract}

\maketitle

\section{Introduction}
Tensor network formalisms have been very successful in describing quantum
many-body systems, and their theoretical study is expected to play a  crucial
role for the  understanding of strongly correlated phenomena in condensed
matter physics. Despite the  huge Hilbert space associated to the many-body
system, scaling exponentially with the
number of particles, the introduction of the density matrix renormalization
group \cite{whiteprl69,schollwoeck} and matrix product states algorithms to
simulate ground states of one dimensional quantum many-body systems
\cite{vidalprl93,verstraeteprl93} have given strong evidence for the fact
that physically interesting states are confined to reside in a small
submanifold of the full Hilbert space. This observation was later proven in
the context of quantum information theory in terms of entanglement properties
of ground states. More specifically, it was shown that ground states of one
dimensional systems whose Hamiltonian is gapped are only weakly entangled
(they obey the area law) and can as such be faithfully and efficiently simulated in terms of matrix product states \cite{faith}. 
The main point of the tensor network formalism is that many-body quantum
states are described in terms of local tensors (or matrices) where the number
of parameters scales polynomially with the system size, and this in turn
makes the classical simulations tractable. The matrix product state formalism
was later generalized to two spatial dimensions by the introduction of 2
different methods, the  multiscale entanglement renormalization ansatz (MERA)
\cite{vidalmera,cincio,evenbly} and the  projected entangled pair states (PEPS) \cite{peps,murg} both of which have later also been extended to fermionic systems \cite{kraus,carlos,corboz0,corboz,barthel,pizornfpeps,zhou}.
Tensor network methods have a great potential in describing two
dimensional quantum systems as they do not suffer from the notorious ``sign
problem'' which makes frustrated spin systems and fermionic spin systems
essentially untractable by quantum monte carlo methods. However, both MERA and
PEPS suffer from relatively high computational complexity. The performance of
PEPS algorithms is furthermore hindered by the instabilities that arise due to
the lack of a normal form of the PEPS structure; this has to be opposed to the
optimization using matrix product states in one dimension  where such a normal form makes the environment essentially disappear from the local optimization procedure.
 
 The environment plays an important role in the context of both density matrix
 renormalization group and tensor network methods as it is responsible for a
 faithful projection of quantum states to a bounded submanifold of the Hilbert
 space, technically known as truncation. Still, the complexity of the PEPS algorithm
 scales as $O(D^{12})$ for open boundary conditions which essentially restricts the computations to small bond dimensions of $D \leq 5$ for finite size PEPS algorithms 
 (but $D \leq 8$ for infinite PEPS algorithms) 
 which is much smaller than in the one dimensional systems where one can reach
 bond dimensions of a few thousands. Note however that that much smaller bond
 dimensions should give already reasonable results; this follows from the
 monogamy properties of entanglement. Another crucial difficulty in simulating
 two dimensional many-body systems in terms of PEPS is however not only the computational complexity itself but also instabilities which occur due to the lack of the normal PEPS structure as is the case in one-dimensional systems. It turns out that the objects in the PEPS algorithm become less and less conditioned with increasing bond dimension and cutting ill-defined components immediately induces effective reduction of the bond dimension. 
 This requires drastic changes to the PEPS simulation techniques, not so much in terms of complexity but rather in terms of stability, calling for the elimination of the double layer structure which is the root of both stability and complexity issues. 
The first step in this direction was made by not calculating the environment
at all but letting it evolve in the process of imaginary time evolution
\cite{xiang}. Such an approach allows to achieve very large bond dimensions
 \cite{ling} but nevertheless seems to require a good initial approximation and works best for translation invariant PEPS states.

In this paper we provide answers to both questions concerning the  complexity and stability of the PEPS algorithms.
We propose a method to simulate the time evolution of PEPS state using an
approximate effective environment, avoiding manipulations with the double
layer tensor network. The approach is justified by the fact that the effective
environment indeed exists and reproduces the effect of the full environment on
the system exactly. Similarly as in the density matrix renormalization group
method, the environment is approximated by a system of particles of an
effective physical dimension which faithfully describe the effect of the
environment to the system. The approach allows for high bond dimensions $D=10$
or more, although the calculation of energy and other observables still
requires the full calculation with the double layer tensor network
structure. Alternatively, quantum monte carlo sampling can be  used to calculate the energy and expectation values of tensor networks with a large bond dimension \cite{schuchqmc,sandvikvidal,ling}.

\section{Method}
A projected entangled pair state (PEPS) on a rectangular $m\times n$ lattice of qubits is parametrized
in terms of local tensors $\underline{A}^{[i,j]\, s_{i,j}}$ for sites $(i,j) \in \{1,\ldots,m\}\times\{1,\ldots,n\}$ with local physical configurations $s_{i,j}$ as
\begin{equation}
\ket{\Psi} = \sum_{\{s_{i,j}\}}\Tr\Big( \underline{A}^{[1,1]s_{1,1}}  \cdots 
\underline{A}^{[m,n]s_{m,n}} \Big) 
\ket{s_{1,1} \cdots s_{m,n} }
\label{eq:PEPS}
\end{equation}
where tensors are contracted along the corresponding horizontal and vertical bonds between neighboring sites. The bonds connecting tensors across horizontal and vertical bonds between the sites are of dimension $D$ such that $\underline{A}^{[i,j]\, s_{i,j}} \in \mathbb{C}^{D\times D\times D\times D}$. The symbol $\Tr$ denotes the ``tensorial trace''  and implies contraction along the boundaries of the square. We shall assume open boundary conditions such that $\Tr$ will only refer to a map from a high-rank tensorial object (of unit size) resulting from the contraction of the square, to a scalar.

Let us consider a bipartite splitting of the $m\times n$ system of qubits to a part consisting of a contiguous block of $M$ qubits and a part containing the remaining $(mn-M)$  qubits. The first part we shall call the subsystem \textit{S} and the second part the environment \textit{E}.
If the subsystem \textit{S} is subject to a local transformation resulting from e.g. a Suzuki-Trotter decomposition of the evolution operator, the internal bond dimension between the qubits in the subsystem \textit{S} will increase. In order to keep the tensor network description~(\ref{eq:PEPS}) manageable, the tensors $\underline{A}^{[i,j]\, s_{i,j}}$ for $(i,j) \in \textit{S}$ must be truncated such that no bond dimension exceeds the chosen bond dimension $D$. This is where the environment \textit{E} comes into play due to the entanglement with the subsystem \textit{S}.

In this section we shall first show how the environment can be efficiently approximated by an effective environment which has approximately the same effect to the subsystem \textit{S}. In the following we shall use the effective environment to manipulate the subsystem \textit{S} to lower the internal bond dimension. This is the core element in the temporal evolution of PEPS states where the evolution operator is decomposed using the Suzuki-Trotter decomposition into a product of local gates. 

For concreteness (and without lose of generality) let us consider a bipartition of the system where the subsystem \textit{S} only contains two neighboring qubits in the horizontal direction, let us call them $\mu \equiv (I,J)$ and $\mu' \equiv (I,J+1)$, which are for simplicity assumed not to be lying on the system boundary. If the subsystem \textit{S} had been subject to a nontrivial local transformation, then the bond dimension between sites $\mu$ and $\mu'$ has increased which calls for a truncation with the help of the environment consisting of all other sites. 
Let us contract the tensor network of the environment tensors 
$\{\underline{A}^{[\nu] s_{\nu}}, \nu \in \textit{E} \}$ which results in a joint environment tensor
$\underline{E}^{[\mu,\mu'] s_{\rm E}}$ where $s_{\rm E} \equiv (s_{\nu}, \nu\in\textit{E})$ denotes the physical (many-body) configuration of the environment sites. The PEPS state~(\ref{eq:PEPS}) is now rewritten to a compact form
\begin{equation}
\ket{\Psi} = \sum_{s,s', s_{\rm E}} \Tr\Big( %
\underline{A}^{[\mu] s}\underline{A}^{[\mu']\, s'} \underline{E}^{[\mu,\mu'], s_{\rm E}}
\Big) %
\ket{s, s'} \ket{s_{\rm E}} 
\label{eq:PEPS1}
\end{equation}
Despite exponentially large physical dimension of the environment, $s_{\rm E} \in \{1,2,\ldots, 2^{mn-2}\}$, it is only connected to the system \textit{S} through six virtual bonds of dimension $D$, 
adding up to a polynomially scaling bond dimension $D^6$. Let us now decompose the state of the whole system to two parts by introducing an over complete set of states spanning the 
subsystem~\textit{S},
\begin{equation}
\ket{\psi^{\rm S}_{(l u u' d d' r)} } = \sum_{s,s',c} A^{[\mu] s}_{l,c,u,d} A^{[\mu'] s'}_{c,r,u',d'} \ket{s,s'}
\label{eq:psiS}
\end{equation}
whereas the environment is written as a superposition of configuration states in the environment as $\ket{\psi^{\rm E}_{(l u u' d d' r)} } = \sum_{s_{\rm E}} E^{[\mu,\mu'] s_{\rm E}}_{l r u u' d d'} \ket{s_{\rm E}}$. The PEPS state~(\ref{eq:PEPS1}) now takes a simple form
\begin{equation}
\ket{\Psi} = \sum_{j=1}^{D^6} \ket{\psi^{S}_j} \ket{\psi^{E}_j} \quad \textrm{with} \quad
j \equiv (l u u' d d' r).
\end{equation}
Due to the entanglement between the subsystem \textit{S} and the environment, the former is in a mixed state given by the reduced density matrix $\rho_S = {\rm tr}_{E} \ket{\Psi}\bra{\Psi}$ which reads (up to a normalization factor)
\begin{equation}
\rho_S \propto \sum_{j,k} \braket{\psi_k^E}{\psi_j^E} \ket{\psi_j^S} \bra{\psi_k^S}
\label{eq:rhoS}
\end{equation}
If one is to apply a local transformation to the subsystem \textit{S}, such as a Trotter gate, the only relevant quantity to consider is the reduced density operator $\rho_S$ which in turn only depends on the environment through inner products $\braket{\psi_k^E}{\psi_j^E}$ and not the state of the environment itself. This leads to the conclusion that, for a fixed set of basis states for the system \textit{S},  there exists an effective environment of physical dimension $D^6$ which exactly reproduces these inner products and is given by
\begin{equation}
\ket{\psi_j^{\rm E}} = \sum_{s_E = 1}^{D^6} {\tilde E}_{j}^{[\mu,\mu'] {\tilde s}_{\rm E}}  \ket{{\tilde s}_{\rm E}}
\end{equation}
where $\underline{\tilde E}^{[\mu,\mu']}$ is obtained by an orthogonal factorization e.g. 
$\mathbf{E} = \mathbf{Q} \mathbf{R}$ where 
$[\mathbf{E} ]_{(s_{\rm E}), (lruu'dd')} \equiv E^{[\mu,\mu'] s_{\rm E}}_{lruu'dd'}$ and 
${\tilde E}_{(lruu'dd')}^{[\mu,\mu'] {\tilde s}_{\rm E}} = [\mathbf{R}]_{({\tilde s}_{\rm E}), (lruu'dd')}$;
the unitary matrix $\mathbf{Q}$ is thus irrelevant.
Such an effective environment suggests that the approximate contraction of the tensor network should be done already on the level of quantum states i.e. in the single layer picture, by truncating not only the virtual but also the physical degrees of freedom of lesser importance.

%%%%%%%%%%%%%%%%%%%%%%%%%%%%%%%%%%%%%%%%%%%%%%%%%%%%%%%%%%%%%%%%%%%%%%
% 	SINGLE LAYER CONTRACTION
%%%%%%%%%%%%%%%%%%%%%%%%%%%%%%%%%%%%%%%%%%%%%%%%%%%%%%%%%%%%%%%%%%%%%%
\subsection{Single layer contraction}
\begin{figure}
\centering
\includegraphics[width=0.99\columnwidth]{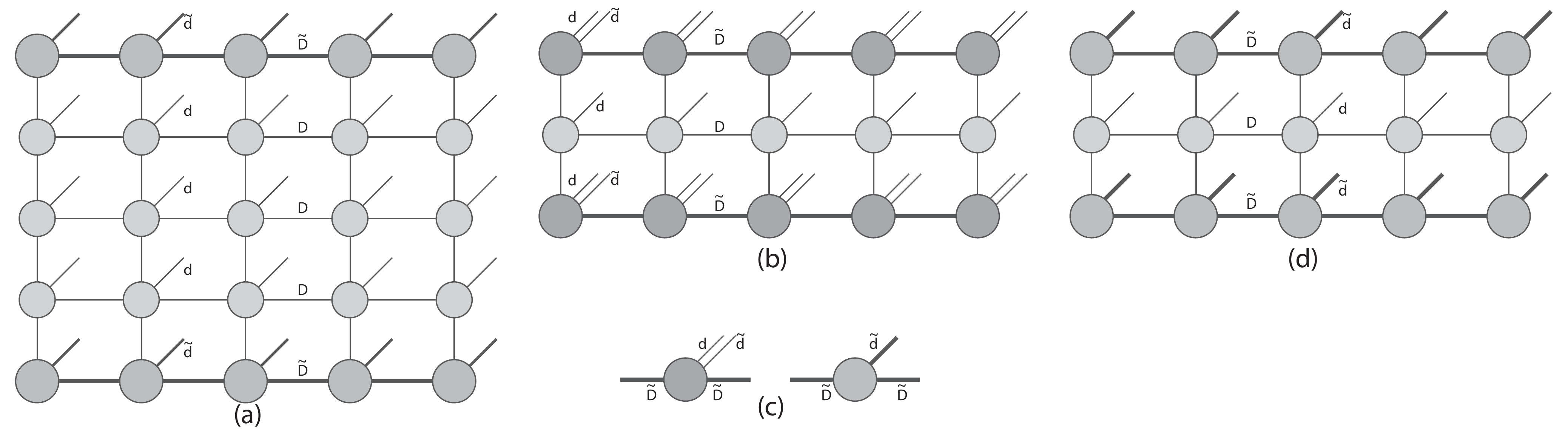}
\caption{Single layer contraction of PEPS structure~(a): contraction of two rows over vertical bonds~(b) is followed by the truncation of the physical bond~(c) resulting in (d).}
\label{fig:singlecontraction}
\end{figure}

Let us show how the effective environment $\underline{\tilde E}^{[\mu,\mu']}$ can be determined efficiently as depicted on Fig.~\ref{fig:singlecontraction}.
The contraction takes place on the single layer with a bond dimension $D$ (as compared to $D^2$ for a double layer structure) and results in an effective environment consisting of particles with a chosen effective physical dimension ${\tilde d}$ connected by virtual bonds of dimension ${\tilde D}$. The tensor network is contracted row by row (or column by column) starting from above and from below such that the $I$-th row is surrounded by a single row of effective particles on both sides.

The first step in the contraction scheme is well known \cite{verstraetempo} in the framework of tensor networks and involves contraction of two rows into a single row with an enlarged horizontal dimension and essentially squared physical bond dimension. In order to make the method efficient, the resulting row must be truncated to a row with a bounded bond dimension ${\tilde D}$ and a bounded physical dimension ${\tilde d}$. 
Let us consider a single layer contraction where rows are merged row-by-row starting from both upper and lower-most row. The result of contracting over vertical virtual bonds connecting two rows results in a matrix product state
\begin{equation}
\ket{\Phi} = \sum_{\underline{s}_j,v_j} {\rm tr}(\mathbf{R}^{[1] \underline{s}_1 v_1 } \cdots \mathbf{R}^{[m] \underline{s}_m v_m }) \ket{\underline{s}_1,v_1,\ldots,\underline{s}_m,v_m}
\label{eq:mpslarge}
\end{equation}
 with a three-fold external bond dimension $(\underline{s_j},v_j) \equiv (s_j,s'_j,v_j)$ where 
 $s_j$ and $s'_j$ are physical bonds at sites $(1,j)$ and $(2,j)$, respectively, and $v_j$ is the vertical bond connecting site $(2,j)$ to site $(3,j)$. The matrix product state~(\ref{eq:mpslarge}) with large matrices $\mathbf{R}^{[j] \underline{s}_j v_j} \in \mathbf{C}^{{\tilde D}D \times {\tilde D} D}$ can easily be approximated by a matrix product state $\ket{\tilde\Phi}$
\begin{equation}
\ket{\tilde \Phi} = \sum_{\underline{s}_j,v_j} {\rm tr}(\mathbf{\tilde R}^{[1] \underline{s}_1 v_1 } \cdots \mathbf{\tilde R}^{[m] \underline{s}_m v_m }) \ket{\underline{s}_1,v_1,\ldots,\underline{s}_m,v_m}
\label{eq:mpssmall}
\end{equation}
 with smaller matrices
 $\mathbf{\tilde R}^{[j] \underline{s}_j v_j} \in \mathbf{C}^{{\tilde D}\times {\tilde D}}$ such that the Euclid distance $\vert\vert \ket{\Phi}-\ket{\tilde\Phi}\vert\vert_2$ is minimal.
 However, the physical bond dimension remains ${\tilde d} d$ instead of the initial ${\tilde d}$ and continuing the procedure would result in an exponentially growing physical bond dimension. Therefore, the physical bond dimension must be truncated as well as depicted on Fig.~\ref{fig:singlecontraction}c.
If the matrix product state~(\ref{eq:mpssmall}) is written in an equilibrated form \cite{vidalprl93}, i.e. such that any given site in a row is connected to unitary environments on both sides with the Schmidt coefficients explicitly given on the corresponding bonds, a very good approximation to the optimal splitting (which is a quartic problem) can be found by truncating the physical dimension at each site $j$ independently by finding matrices 
$\mathbf{\tilde B}^{[j] {\tilde s} v}$ for which 
$\sum_{{\tilde s}} \mathbf{\tilde B}^{[j] {\tilde s} v} \otimes \mathbf{\tilde B}^{[j] {\tilde s} v' *}$ best approximates 
$\sum_{\underline{s}} \mathbf{\tilde R}^{[j] \underline{s} v} \otimes \mathbf{\tilde R}^{[j] \underline{s} v' *}$ according to the Frobenius norm. Such matrices are easily found from the singular value decomposition
$\mathbf{\tilde R}$,
\begin{equation}
{\tilde R}_{l r}^{[j] \underline{s} v} = \sum_{\tilde s} U_{l r}^{[j] v {\tilde s}} \Sigma_{\tilde s}^{[j]} V_{\underline{s} {\tilde s}}^{[j]}, 
\label{eq:svd1}
\end{equation}
as $\mathbf{\tilde B}^{[j] {\tilde s} v} = \mathbf{U}^{[j] v {\tilde s}} \Sigma^{[j]}_{\tilde s}$.
If all singular vectors of $\mathbf{U}$ were retained where $[\mathbf{U}]_{(lrv), s} = U_{lr}^{[j] v s}$, 
such transformation would be exact while a good approximation to the environment is obtained taking ${\tilde d}$ leading singular vectors of $\mathbf{U}$ in the singular value decomposition~(\ref{eq:svd1}).
In the end, the two rows are described as a single matrix product state with a physical dimension ${\tilde d}$ and a vertical external bond dimension ${\tilde D}$ which guarantees bounded matrices and thus makes the algorithm efficient. While such approximation is not strictly optimal, it nevertheless provides a very good approximation in practice. The idea of truncating the physical degrees of freedom is not new but it is intrinsic to e.g. MERA \cite{vidalmera} and also appears in other renormalization algorithms \cite{chen}.

\begin{figure}
\centering
\includegraphics[width=0.9\columnwidth]{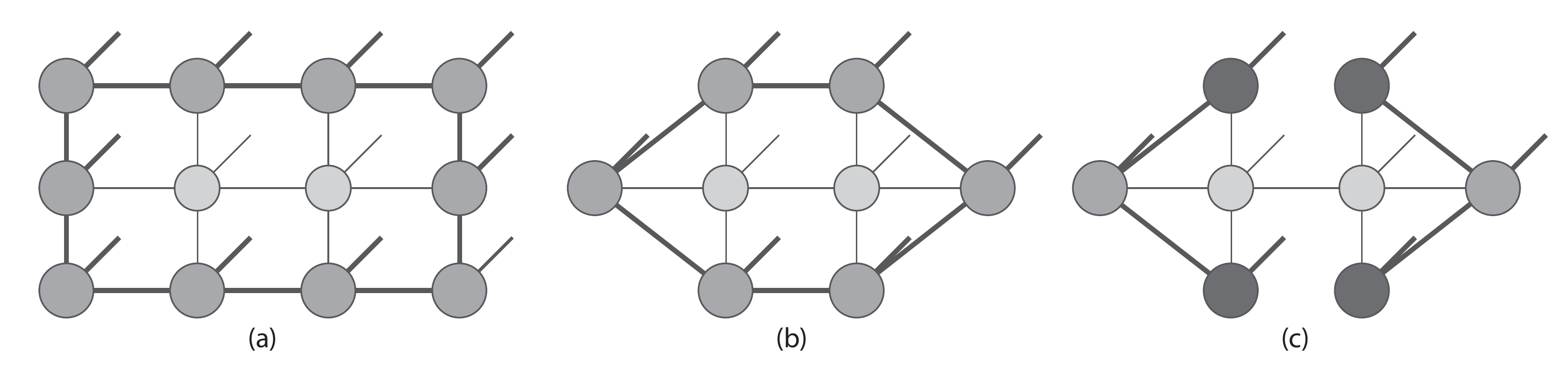}
\caption{Effective PEPS after the single layer contraction (a). Effective environmental sites on corner are absorbed in neighboring sites (b). The environment is approximated by two disconnected parts (c).}
\label{fig:fig3x4}
\end{figure}

The procedure from the previous paragraph is repeated for all rows $i < I$ and $i> I$ starting from the upper ($i=1$)  and the lower ($i=m$) boundary row, respectively, such that the structure depicted on Fig.~\ref{fig:singlecontraction}d is obtained. Eventually, the same procedure is applied in the horizontal direction such that the two sites of interest are surrounded by ten (or less for boundary sites) effective environmental sites as shown on Fig.~\ref{fig:fig3x4}a.
An effective PEPS structure obtained by this procedure (Fig.~\ref{fig:fig3x4}a) can be further simplified by absorbing corner sites to their neighbors (Fig.~\ref{fig:fig3x4}b) which is done in a trivial way followed by an exact reduction of the effective physical bond (Fig.~\ref{fig:singlecontraction}c).

This way an effective environment $\underline{\tilde E}^{[\mu,\mu']\, s_{\rm E}}$ is obtained 
which is in a straight-forward way related to the effective double-layer norm operator 
$\mathcal{N}^{[\mu,\mu']}$ as
\begin{equation}
\underline{\mathcal{N}}^{[\mu,\mu']} 
= \sum_{s_{\rm E}} \underline{E}^{[\mu,\mu'] s_{\rm E} *} \otimes \underline{E}^{[\mu,\mu'] s_{\rm E}}
\label{eq:Neff}
\end{equation}
where its matrix elements represent the inner products $\mathcal{N}^{[\mu,\mu']}_{ij} \equiv \braket{\psi^E_i}{\psi^E_{j}}$ as defined previously. In the context of the usual time evolution, the effective operator $\mathcal{N}$ is used to determine a new set of tensor $\{ A^{[\mu]}, A^{[\mu']} \}$ after a Trotter step has been applied on sites $(\mu,\mu')$.

Similarly to the double layer contraction, the single layer contraction becomes exact for sufficiently large truncation parameters ${\tilde D}$ and ${\tilde d}$ whereas good approximations can be obtained already with ${\tilde d} = {\tilde D} = D$.
On the other hand, this way of calculating the environment is in many ways advantageous to the conventional contraction of the double layer tensor network. The first advantage is the computational cost. While the complexity of the conventional double layer contraction scales as $O(D^{12})$, the costs of the single layer truncation only scale as $O(D^7)$. The second advantage is that by doing the single layer truncation, good estimates can be obtained of how to perform gauge transformations on the original environment sites connected to the subsystem \textit{S}, that the effective environment norm $\mathcal{N}$ becomes better conditioned \cite{inprep}, which is of crucial importance for the stability of the algorithm.

Let us now consider a Trotter gate $T^{[\mu,\mu']}$ acting on two neighboring sites $(\mu,\mu')$
and find a matrix product state ${\tilde \Psi}$ defined as
\begin{equation}
\ket{ {\tilde \Psi} } = \sum_{s,s', s_{\rm E}} \Tr\Big( %
\underline{\tilde A}^{[\mu] s}\underline{\tilde A}^{[\mu']\, s'} \underline{E}^{[\mu,\mu'], s_{\rm E}}
\Big) %
\ket{s, s'} \ket{s_{\rm E}} 
\label{eq:PEPS1x}
\end{equation}
which best approximates $T^{\mu,\mu'} \ket{\Psi}$
such that the bond dimension between the tensors on sites $(\mu,\mu')$ is upper-bounded by $D$.
In the conventional update scheme, the tensors $\{ {\tilde A}^{[\mu]}, {\tilde A}^{[\mu']} \}$ are obtained by solving a multi-quadratic optimization problem in an iterative way which involves solving a linear system of equations 
$N_{\rm eff}^{[\mu]} A^{[\mu]} = b^{\mu}$ where the $N_{\rm eff}^{[\mu]}$ is an effective norm operator for the site $\mu$, obtained by contracting $\mathcal{N}^{[\mu,\mu']}$ and tensors at the site $\mu'$ (and similarly for site $\mu'$) \cite{pepo}. While the gauge transformations allow us to make $N_{\rm eff}^{[\mu]}$ equal to the identity in the case of matrix product states, this is not possible in the case of PEPS states. Furthermore, the linear system of equations in consideration is typically ill-conditioned due to emergence of very small eigenvalues in the effective norm operator $N_{\rm eff}$ leading to instabilities, especially for large bond dimensions $D=3,4$, which become more and more pronounced in the process of simulation as we shall show later. 
In the usual PEPS time evolution, this problem is evaded by the projection to the ``well'' defined subspace spanned by the singular vectors of $N_{\rm eff}$ with respect to a cut-off parameter $\varepsilon$ determining the ratio between the smallest kept and the maximal singular value.
If the parameter is set too high, the effective bond dimension is largely reduced whereas in the case of too low setting some ill-conditioned components are also retained resulting in instabilities.
 While one could use $E^{[\mu,\mu']}$ obtained by the single layer contraction to calculate the effective norm operator~(\ref{eq:Neff}) and then use the conventional update scheme, that would still not solve the stability issues. 

%%%%%%%%%%%%%%%%%%%%%%%%%%%%%%%%%%%%%%%%%%%%%%%%%%%%%%%%%%%%%%%%%%%%%%
% 	HOW TO CUT THE ENVIRONMENT INTO TWO HALVES
%%%%%%%%%%%%%%%%%%%%%%%%%%%%%%%%%%%%%%%%%%%%%%%%%%%%%%%%%%%%%%%%%%%%%%
\subsection{Environment splitting}
In order to eliminate the stability problems, we will show how to avoid calculations with the double layer tensors such as $\mathcal{N}$ defined~(\ref{eq:Neff}). The core of the problems is that the 
environment appears as a cyclic matrix product operator (see Fig.~\ref{fig:fig3x4}b) and as such does not permit the standard way of finding the optimal tensors 
$\{ \underline{\tilde A}^{[\mu] s }, \underline{\tilde A}^{[\mu'] s } \}$ using the singular value decomposition. We however know empirically, that the environment of two sites in sufficiently large lattices can be fairly well approximated by a product of two separate environments (Fig.~\ref{fig:fig3x4}c) which is the idea we will pursue in the following.

\begin{figure}
\includegraphics[width=0.9\columnwidth]{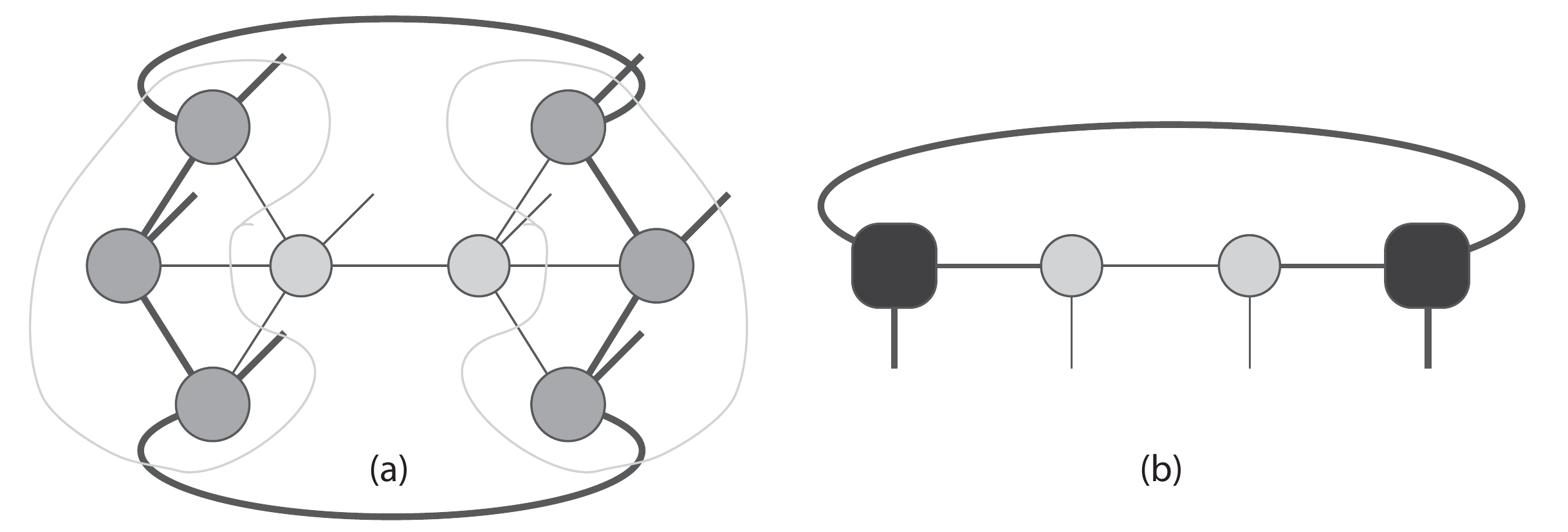}
\caption{A PEPS state with an effective environment (a) (identical to Fig.~\ref{fig:fig3x4}b) can be understood as a matrix product state with periodic boundaries (b).}
\label{fig:peps2mps}
\end{figure}
Let us rewrite the PEPS state~(\ref{eq:PEPS1}) as a matrix product state with periodic boundaries (see Fig.~\ref{fig:peps2mps}) as
\begin{equation}
\ket{\Psi} = \sum_{\underline{s}} 
{\rm tr} 
\Big( %
\mathbf{L}^{s_L}
\mathbf{A}^{[\mu] s}
\mathbf{A}^{[\mu'] s'}
\mathbf{R}^{s_R}
\Big) %
\ket{s_L,s,s',s_R}
\label{eq:mpspbc}
\end{equation}
where $[\mathbf{A}^{[\mu] s} ]_{(lud) r} = A^{[\mu] s}_{lrud}$, 
$[\mathbf{A}^{[\mu'] s} ]_{l (udr)} = A^{[\mu'] s}_{lrud}$ whereas 
$\mathbf{L}^{s}$ and $\mathbf{R}^{s}$ correspond to the contraction of tensors belonging to the left and right three sites on Fig.~\ref{fig:fig3x4}b. 

There is no way known to us how to found the optimal matrices $\mathbf{\tilde A}^{[\mu] s}$ and 
$\mathbf{A}^{[\mu'] s'}$ exactly without employing the sweeping optimization mechanism described in the previous section. However, there are several ways to split the environment approximately, assuming that the internal correlations between the two parts of the environment decay sufficiently fast. 
The first possibility which gives remarkably good results is to do the singular value decomposition of both parts of the environment,
\begin{equation}
L^{\gamma, s_L, l} = \sum_{\tilde s} U^{[L]}_{(\gamma,s),{\tilde s}} \Sigma_{\tilde s}^{[L]} V_{l, {\tilde s}}^{[L] *}
\end{equation}
and then taking the left approximate environment as
\begin{equation}
{\tilde L}^{ {\tilde s}}_{l} = \Sigma_{\tilde s}^{[L]} V_{l, {\tilde s}}^{[L] *}.
\end{equation}
The right part of the environment is transformed in a similar way. In practice, the singular value is done separately for values of the internal environment bond $\gamma$, followed by the singular value decomposition of the concatenated and weighted right singular vectors which is numerically favorable.

The approach in the previous paragraph is rather expensive in our case due to the large physical dimension of the environment sites (${\tilde d}^3$). For that reason, we will pursue in an even simpler way by simply self-contracting the internal environment bond for each part of the environment
\begin{equation}
{\tilde L}^{ {\tilde s}}_{l} = \sum_{\gamma} L^{s}_{\gamma, l}, \quad
{\tilde R}^{ {\tilde s}}_{r} = \sum_{\gamma} R^{s}_{r, \gamma}.
\end{equation}
The result is a single matrix product state with open boundary conditions for which the optimal matrices $\mathbf{\tilde A}^{[\mu] s}$ and $\mathbf{\tilde A}^{[\mu'] s'}$ are determined exactly by the singular value decomposition. 
From numerical tests we observe that this approach is only slightly less accurate than the one presented earlier. The total cost of this step scales as $O(D^9)$, all coming from the SVD, but it is in practice negligible for relatively small dimensions $D\sim 10$ when compared to the single layer contraction part of the method involving fairly many steps scaling as $O(D^7)$.

\section{Results}
To illustrate the validity of the method we consider an antiferromagnetic Heisenberg model on a square lattice
\begin{equation}
H = \sum_{ \langle \mu\nu \rangle} \big(\sigma_{\mu}^{\rm x} \sigma_{\nu}^{\rm x} + 
\sigma_{\mu}^{\rm y} \sigma_{\nu}^{\rm y} + 
\sigma_{\mu}^{\rm z} \sigma_{\nu}^{\rm z}\big)
\label{eq:H}
\end{equation}
for which the ground state properties are accurately described by the stochastic series expansion (SSE) quantum monte carlo method \cite{sandvik}. 
We perform the imaginary time evolution $\ket{\Psi(\beta)} = e^{-\beta H} \ket{\Psi_0}$ using the second order Suzuki-Trotter expansion where two-site local gates are applied to the PEPS state, followed by the truncation of the corresponding tensors. 

\begin{figure}
\centering
\includegraphics[width=0.99\columnwidth]{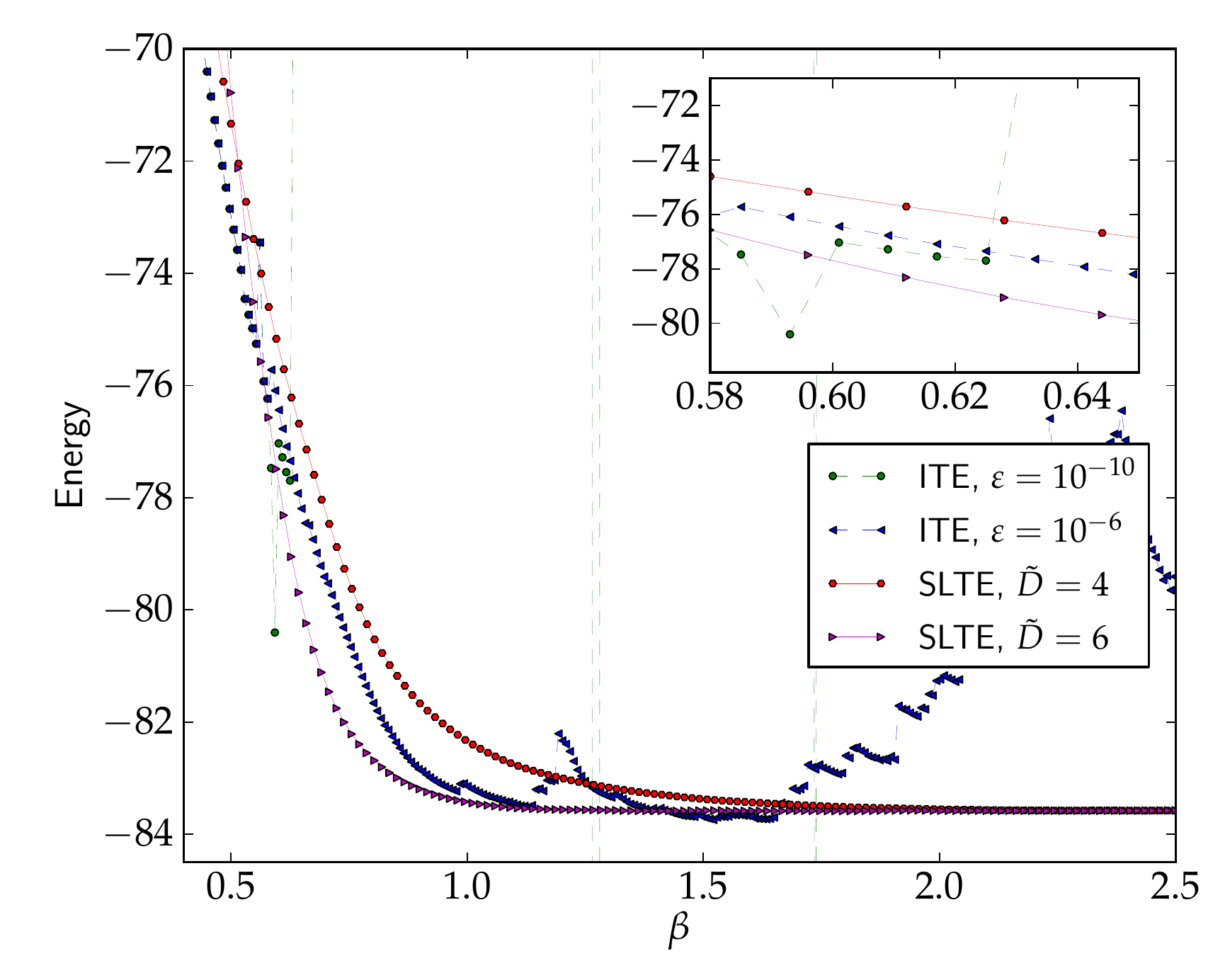}
\caption{Imaginary time evolution of a random initial state with bond dimension $D=2$ under Hamiltonian~(\ref{eq:H}) on a $6\times 6$ system using the usual PEPS time evolution algorithm (ITE) and the single layer time PEPS algorithm (SLTE). The Trotter time step was set to $10^{-3}$. The cutoff in the linear problem $\varepsilon$ appearing in ITE is designated in the legends. The ${\tilde D}={\tilde d}$ for the SLTE is given in the legends whereas ${\tilde D}=64$ for the ITE. The exact ground state energy equals  $E_0 = -86.9$.}
\label{fig:plotinstab}
\end{figure}
First we compare the proposed method to the usual imaginary time evolution of PEPS \cite{murg} with a bond dimension $D=2$ on a system of $6\times 6$ qubits.
In the single layer method we choose ${\tilde D} = {\tilde d} = 4, 6$ whereas the cut-off parameter in the usual PEPS time evolution is set to $D_{\rm cut} = 64$. In the latter, ill-conditioned linear systems require another cut-off parameter which is chosen as $\varepsilon = 10^{-10}, 10^{-6}$.
The results in Fig.~\ref{fig:plotinstab} confirm that the usual time evolution results in instabilities due to the ill-conditioned linear problems which are being solved in the simulation. Increasing the cut-off parameter $\varepsilon$ from $10^{-10}$ to $10^{-6}$ suppresses the non-physical solutions and pushes the simulation forward, although the accuracy of simulation steps is reduced. Since the determination of the cut-off parameter is heuristic procedure and inevitably results in either cutting relevant degrees of freedom or keeping non-physical ones, the simulation eventually becomes unstable.
This phenomenon is even more pronounced with larger bond dimensions where the linear problems are of larger dimension and it is even more difficult to make a sensible compromise for $\varepsilon$.
From the technical point of view, the PEPS tensors are always rescaled such that their $2$-norm is the same for all tensors. Increasing the cutoff parameter ${\tilde D}$ to $128$ did not help significantly for $\epsilon=10^{-10}$.

The single layer time evolution, on the other hand, produces no instabilities for arbitrarily long times, although the results do oscillate slightly (not noticeable on the figure). Choosing a larger effective bond dimensions ${\tilde D}$ and ${\tilde d}$ makes the convergence faster but nevertheless results in an comparably good final state. We note, however, that the single layer time evolution involves approximations and slightly more accurate results can be achieved (note a few points for ITE, $\varepsilon=10^{-6}$) using the usual time evolution.
The time needed to obtain the results by the usual time evolution is by a factor of hundred larger than the single layer time evolution. Both approaches were done using the same equally (non-)optimized code  and starting from the same random initial state. For smaller bond dimensions, the single layer approach can thus be used to quickly obtain a very good initial approximation which could be further refined by some double layer technique such as minimizing the energy by sweeping \cite{peps}.

Secondly, we test the method on larger systems and larger bond dimensions. 
For a given bond dimension, the simulation was running as long as the energy decay rate was sufficiently high and the results were used as the initial state for simulations with a larger bond dimension $D+1$. The growth from $D$ to $D+1$ is intrinsic to the time evolution and does not require any zero-padding.

\begin{figure}
\centering
\includegraphics[width=0.99\columnwidth]{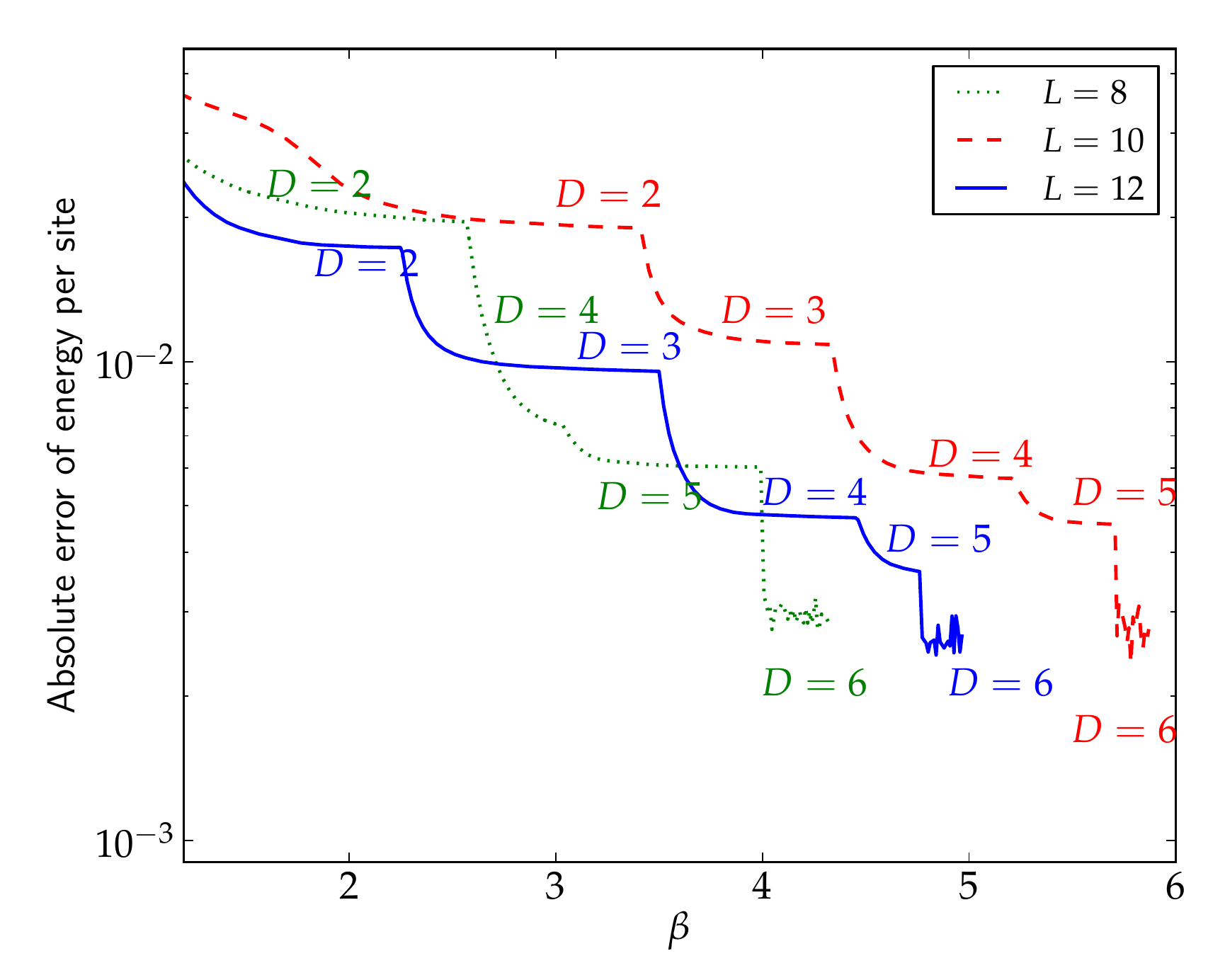}
\caption{Absolute error of the energy per site for the antiferromagnetic Heisenberg model~(\ref{eq:H}) as a function of iteration step and lattice sizes $L\times L$ as given in the legend. The results for $D=6$ were obtained by sampling.}
\label{fig:plotenergy}
\end{figure}
In Fig.~\ref{fig:plotenergy} we present the results for the absolute error of the energy per site for the antiferromagnetic Heisenberg model~(\ref{eq:H}) compared to the results obtained by the stochastic series expansion which we take as exact. We consider three system sizes and bond dimensions up to $D=6$. The single layer truncation parameters were in all cases chosen as ${\tilde D} = {\tilde d} = D$. Note that this model is critical and is among fairly difficult models to simulate using tensor network methods.
From the results for $D \leq 5$ where the energy is calculated by the approximate contraction of the PEPS tensor network, it is clearly visible that the energy decreases monotonically until the plateau is reached which justifies the single layer contraction scheme and the approximate environment splitting.

We can easily simulate PEPS systems of $10\times 10$ sites with bond dimensions $D=10$ or even more, however the extraction of expectation values such as the energy is nontrivial unless we also transform the hamiltonian itself by the isometries generated in the single layer contraction procedure as known in the context of the DMRG. For purposes of this manuscript we rely on the double layer contraction scheme to calculate the energy which, much more than in the translation invariant infinite PEPS algorithm, is computationally very costly and at present only sensible for bond dimensions $D \sim 5$. For that reason, the energies of the PEPS with a large bond dimension $D=6$ are calculated by sampling using the Metropolis algorithm following the fact \cite{schuchqmc,sandvikvidal} that
\begin{equation}
E = \frac{ \sum_{\mu} p_{\mu}  \frac{ \bra{\mu} H \ket{\Psi} }{ \braket{\mu}{\Psi}  }   } { \sum_{\mu} p_{\mu} } 
\quad \textrm{where} \quad
p_{\mu} = \vert \braket{\mu}{\Psi} \vert^2
\end{equation}
Metropolis algorithm also allows us to sample not $\ket{\Psi}$ but rather $P\vert_{S^{\rm z}=0} \ket{\Psi}$ where $P\vert_{S^{\rm z}=0}$ is a projection operator to the zero total spin subsector.
The benefit of the Metropolis sampling is again the single layer picture of the problem which avoids doubling the virtual bond dimension and thus reduces the contraction cost.
The downside however is that one has to carefully choose the Metropolis updates otherwise the variance decays relatively slowly as visible from Fig.~\ref{fig:plotenergy} where the variance is still large. However, the results obtained by sampling for $D=6$ are consistent with the results for $D=5$ obtained by the approximate contraction of the tensor network.

\section{Conclusion}
We have presented an approximate method to simulate quantum many-body systems on finite two-dimensional lattices using the projected entangled pair states (PEPS) where the contractions are only done on the level of quantum states (single layer tensor network).  This contrasts with the conventional PEPS simulation scheme where the simulations involve contractions on the level of expectation values given by a double layer tensor network with a squared bond dimension, leading to a high computational cost. The single layer approach eliminates the stability issues present in both the time evolution of PEPS and the variational PEPS algorithm to simulate ground states, which in both cases originate from the ill-conditioned double layer effective norm operator. Unlike the usual time evolution which is exact for sufficiently large cut-off parameters and arbitrary precision arithmetics, the single-layer time evolution is approximate due to its simplified treatment of the environment.
We compared the single layer time evolution to the usual time evolution for PEPS and observed comparable results in accuracy whereas the stability of the simulation is better in the single layer approach.
We tested the method for larger bond dimensions on an isotropic antiferromagnetic Heisenberg model on a square lattice where we again observed monotonic decrease of the energy for bond dimensions $D \leq 5$ where it can be calculated by contracting the PEPS approximately. We presented sampled results for $D=6$ which are consistent with the results for lower bond dimensions.
While we only presented the results for a conceptually simple model, the method can easily be applied to any spin or fermionic system in two spatial dimensions. The extension of the single layer contraction technique to the latter case is straight-forward where all arising sign factors can be absorbed locally due to the well defined parity of tensors of the fermionic PEPS~\cite{kraus}.

\begin{acknowledgments}
This work was supported by the FWF grants FoQuS and ViCoM,  and the ERC grant QUERG.
The computational results presented have been achieved using the Vienna Scientific Cluster (VSC).
\end{acknowledgments}


\begin{thebibliography}{10}

\bibitem{whiteprl69}
S. R. White, Phys. Rev. Lett. \textbf{69}, 2863 (1992).
    
\bibitem{schollwoeck}
U. Schollw\"{o}ck, Rev. Mod. Phys. \textbf{77}, 259  (2005).

\bibitem{vidalprl93}
G. Vidal, Phys. Rev. Lett. \textbf{93}, 040502  (2004).

\bibitem{verstraeteprl93}
F. Verstraete, D. Porras, and J. I. Cirac, Phys. Rev. Lett. \textbf{93}, 227

\bibitem{faith}
F. Verstraete and J. I. Cirac, Phys. Rev. B \textbf{73}, 094423 (2006).
    
\bibitem{vidalmera}
G. Vidal, Phys. Rev. Lett. \textbf{99}, 220405 (2007).

\bibitem{cincio}
L. Cincio, J. Dziarmaga, M. M. Rams, 
Phys. Rev. Lett. \textbf{100}, 240603 (2008).

\bibitem{evenbly}
G. Evenbly and G. Vidal. Phys. Rev. Lett. \textbf{102}, 180406
(2009).
    
\bibitem{peps}
J. I. Cirac and F. Verstraete, \texttt{arXiv:cond-mat/0407066}.
    
\bibitem{murg}
F. Verstraete, V. Murg, and J. I. Cirac, 
Adv. Phys. \textbf{57}, 143 (2008).


\bibitem{kraus}
C. V. Kraus, N. Schuch, F. Verstraete, and J. I. Cirac, 
Phys. Rev. A \textbf{81}, 052338 (2010).


\bibitem{zhou}
Q.-Q. Shi, S.-H. Li, J.-H. Zhao, H.-Q. Zhou,  \texttt{arXiv:0907.5520}.

\bibitem{corboz0}
P. Corboz and G. Vidal,
Phys. Rev. B \textbf{80}, 165129 (2009).

\bibitem{carlos}
C. Pineda, T. Barthel, and J. Eisert,
Phys. Rev. A \textbf{81}, 050303(R) (2010).

\bibitem{corboz}
P. Corboz, R. Orus, B. Bauer, and G. Vidal, 
Phys. Rev. B \textbf{81}, 165104 (2010).

\bibitem{barthel}
T. Barthel, C. Pineda, and J. Eisert,
Phys. Rev. A \textbf{80}, 042333 (2009).


\bibitem{pizornfpeps}
I. Pi\v{z}orn and F. Verstraete, Phys. Rev. B \textbf{81}, 245110 (2010).


\bibitem{xiang}
H.-C. Jiang, Z. Y. Weng, and T. Xiang, Phys. Rev. Lett. \textbf{101}, 090603 (2008).


\bibitem{ling}
L. Wang, I. Pi\v{z}orn, and F. Verstraete, \texttt{arXiv:1010.5450}.



\bibitem{schuchqmc}
N. Schuch, M. M. Wolf, F. Verstraete, and J. I. Cirac, Phys. Rev. Lett. \textbf{100}, 040501 (2008).

\bibitem{sandvikvidal}
A. W. Sandvik and G. Vidal, Phys. Rev. Lett. \textbf{99}, 220602 (2007).



\bibitem{verstraetempo}
F. Verstraete, J. J. Garc\'ia-Ripoll, and J. I. Cirac, Phys. Rev. Lett. \textbf{93}, 207204 (2004).


\bibitem{chen}
X.~Chen, Z.-C. Gu, and X.-G. Wen,
Phys. Rev. B \textbf{82}, 155138 (2010).


\bibitem{sandvik}
A. W. Sandvik, Phys. Rev. B \textbf{56}, 11678 (1997).



\bibitem{inprep}
I Pi\v{z}orn \textit{et al}, \textit{in preparation}.


\bibitem{pepo}
In practice, the usual time evolution is done by truncating a PEPS obtained by applying a product of all commuting Trotter gates, not just one Trotter gate; the concept is however the same.



\end{thebibliography}
\end{document}